\documentstyle[12pt]{article}
\begin{document}
\title{Double beta decay and the proton-neutron residual interaction}

 \author{Jorge G. Hirsch$^1$\thanks{e-mail: hirsch@fis.cinvestav.mx},
	 Peter O. Hess$^2$\thanks{e-mail: hess@roxanne.nuclecu.unam.mx} and
	 Osvaldo Civitarese$^3$\thanks{Fellow of the CONICET, Argentina;
	e-mail: civitare@venus.fisica.unlp.edu.ar}\\
{\small\it $^1$Departamento de F\'{\i}sica, Centro de Investigaci\'on
y de Estudios Avanzados del IPN,}\\
{\small\it Apdo. Postal 14-740 M\'exico 07000 D.F.}\\
 {\small\it $^2$Instituto de Ciencias Nucleares, Universidad Nacional
Aut\'onoma de M\'exico,}\\
{\small\it Apdo. Postal 70-543, M\'exico 04510 D.F.}\\
{\small\it $^3$ Departamento de F\'{\i}sica, Universidad Nacional de
La Plata, }\\ {\small\it c.c. 67 1900, La Plata, Argentina.}
}
\maketitle

\begin{abstract}
The validity of the pn-QRPA and -RQRPA descriptions of double beta decay
transition amplitudes is analyzed by using an exactly solvable model.
It is shown that the collapse of the QRPA is physically meaningful
and that it is associated with the appearance
of a state with zero energy in the spectrum. It is shown that in the
RQRPA this particular feature is not present and that this approach 
leads to finite but otherwise spurious
results for the double beta decay transition amplitudes
near the point of collapse.
 
\end{abstract}

\noindent
PACS number(s): 21.60.Jz, 21.60.Fw, 23.40.Hc

\vskip .5cm

The neutrinoless double beta decay ($\beta\beta_{0\nu}$) violates
lepton number conservation and requires the existence of
massive Majorana neutrinos \cite{Ver86}. Due to this fact the detection
of this decay has  attracted considerable experimental
effort \cite{Moe93}. The two neutrino mode of the double beta
decay ($\beta\beta_{2\nu}$), on the other hand, can be described as
a second order process in the standard electroweak theory, 
its decay rate being independent of any new physics beyond the
standard model. Both to predict
and to analyze the data on double beta decay require
a precise calculation of the various nuclear matrix elements
needed to compute the corresponding half-lives.

The quasiparticle random phase approximation (QRPA), including a
particle - particle channel in the residual interaction, can reproduce the
experimentally determined two-neutrino double-beta decay
($\beta\beta_{2\nu}$) half-lives. This is so because the transition
amplitudes are strongly suppressed
for certain values of the force parameters. This
result was tested by comparing the results of the QRPA against an
 exactly solvable
model \cite{Vog86,Eng87} and including realistic residual
interactions \cite{Civ87,Mut89}.

However, the predictive power of the QRPA is strongly diminished
because
the ground state to ground state $\beta\beta_{2\nu}$ transition
amplitudes are extremely 
dependent upon the structure of the adopted proton-neutron interaction.
For some critical values of the model parameters the QRPA collapses,
i.e; the energy of the first excited QRPA state vanishes.
It makes the theory nearly useless for some
particular cases, the worst of which is the $\beta\beta_{2\nu}$
decay of $^{100}Mo$ \cite{Vog86,Eng87,Civ91,Civ91b,Gri92}.

A renormalized version of the QRPA (RQRPA) \cite{Har64,Row68}, which
includes some corrections beyond the quasiboson approximation, has
been reformulated recently
\cite{Cat94} and applied to the $\beta\beta_{2\nu}$ decay
\cite{Toi95}. Contrary to the QRPA in the RQRPA there is no
collapse for any 
set of values of the force's parameters. It was presented as a reliable
tool and was applied to the  $\beta\beta_{2\nu}$ decay of
$^{100}Mo$ \cite{Toi95}. Similar results were found with the
inclusion of proton-neutron pairing interactions \cite{Sch96}.

There are some concerns about the formulation of the RQRPA.
The most important one is that it makes use of the exact
commutation relations between phonons, adding one-quasiparticle 
scattering terms to the ordinary quasi-boson approximation,
but it does not include similar terms in the
hamiltonian and in the transition operators. In the present letter it will 
be shown that the collapse of the QRPA correlates
with the presence of a state with zero energy which is not found 
in the RQRPA spectrum.
This shortcoming is not only a
failure of the RQRPA but it is shared by others higher order approximations.
Spurious excitations are exactly separated in the RPA but higher order
approximations can introduce spurious components which finally
dominate the results \cite{Row70}.

The model hamiltonian \cite{Kuz88,Mut92,Civ94a} consists 
of a single
 particle term, a pairing term for protons and neutrons and a schematic
 charge-dependent  residual interaction including particle-hole
 and particle-particle channels.
 It has been shown that this
 interaction, when treated in the framework of the QRPA,
 produces similar results as those obtained by using 
a  G-matrix constructed from the OBEP
Bonn potential in reproducing single- and double-beta decay matrix
elements \cite{Civ94a,Civ94b,Civ95}. In this letter we will consider
 the single shell
limit ($j_p = j_n = j$) and monopole term ($J=0$) of the residual interaction.
As we shall show later on this model, which is not
intended to accurately reproduce actual data, does indeed display 
the qualitative
features of a realistic pn-QRPA calculation. Excitation energies,
single- and double-beta decay transition amplitudes and ground state
correlations, in this model, depend on the particle-particle strength parameter
in the same way as they do in more elaborated calculations with many single
particle levels and with more realistic interactions. 
The present case corresponds to beta decay transitions of the Fermi type.

The schematic hamiltonian reads
\begin{equation}
H = \sum\limits_p e_p a^\dagger_p a_p - G_p S^\dagger_p S_p +
 \sum\limits_n e_n a^\dagger_n a_n - G_n S^\dagger_n S_n +
2 \chi \beta^- \cdot \beta^+
	  - 2 \kappa P^- \cdot P^+ , \label{hamex}
\end{equation}

\noindent with
\begin{equation}
\begin{array}{c}
S^\dagger_p = \sum\limits_p a^\dagger_p a^\dagger_{\bar p}/2 ,\hspace{1cm}
S^\dagger_n = \sum\limits_n a^\dagger_n a^\dagger_{\bar n}/2 ,
\\
 \beta^- = \sum_{i,j} <i|\tau^- |j> a^{\dagger}_i a_j
 \hspace{1cm}
P^- = \sum_{i,j} <i|\tau^- |j> a^{\dagger}_i a^{\dagger}_{\bar j}
\end{array}
\end{equation}

\noindent
$ a^{\dagger}_p = a^\dagger_{j_p m_p}$ being the particle creation
 operator,
 $a^{\dagger}_{\bar p} = (-1)^{j_p -m_p} a^{\dagger}_{j_p -m_p}$ its time
reversal and $\tau^-$ the isospin lowering operator ($\tau^-|n>=|p>$). The 
parameters $\chi$ and $\kappa$ play the role 
of the renormalization factors $g_{ph}$ and $g_{pp}$ introduced in the
literature \cite{Vog86,Eng87,Civ87,Mut89}.

The hamiltonian (\ref{hamex}) can be expressed in terms of
generators of an SO(5) algebra \cite{Par65,Hec65,Kle91}. It can be
reduced to an SO(5) and isospin scalar \cite{Hir96b} if its parameters are
selected as

\begin{equation}
e_p =e_n, \hspace{1cm}\chi=0,\hspace{1cm}G_p=G_n=4\kappa.
\end{equation}

This SO(5) scalar limit was used to study the ``pairing plus monopole
model'' many years ago \cite{Aga68,Sch68}.
 If $\chi \neq 0$ the isospin symmetry is broken in the particle-hole
channel and for $4\kappa \neq G_p (G_n)$ this symmetry is broken in the
particle-particle channel.
 The Hilbert space is constructed using the eigenstates of the
isoscalar hamiltonian, which are labeled by the number of
particles $N$, the isospin $T$ and its projection $T_z$. If $G\ne
4\kappa$ the eigenstates have definite $N$ and $T_z$ but
not good isospin since the Hamiltonian mixes states
with different isospin $T$.

We have selected $N_n > N_p $ and a large $j$ to  mimic the
realistic situation in medium  and heavy nuclei. In this letter we
will use the following parameters:
\begin{equation}
\begin{array}{l}
j = 19/2, \hspace{.5cm} N = 20,\hspace{.5cm} 1 \le T_z \le 5,\\
e_p = 0.7 MeV, \hspace{.5cm} e_n = 0.0 MeV, \hspace{.5cm} G_p = G_n
= 0.2 MeV, \hspace{0.5cm}\chi= 0.025 MeV~.
\end{array}
\end{equation}
\noindent  $\kappa$ will be used as a running parameter.

{\bf Fig. 1}

Lowest energy $0^+$ states of different nuclei  are shown in Fig.
1, for $G= 4\kappa$, in an energy vs. $Z$ plot. States are labeled
by $(T,T_z)$. It is clear that a simple model with both like particles and
pn- pairing interactions can reproduce the qualitative form of the mass parabola.
Fermi  transitions ($\beta^- =  t^-$ ) are allowed between
members of the same isospin multiplet. In our single shell example,
lowest $0^+$ states in each odd-odd nuclei (N-1, Z+1, A) are the isobaric 
analog states corresponding to the states of
the even-even nuclei with (N, Z, A) nucleons.
Ground states of the initial (N, Z, A) and final (N-2, Z+2, A)
even-even nuclei have different isospin and
the Fermi-double-beta  decay is forbidden in the isoscalar limit.

{\bf Fig. 2}

The results shown by a full  line, in Fig. 2, represent the excitation energy
$E_{exc}$ of the lowest $0^+$ state in the odd-odd intermediate
nucleus ($T_z = 3$) with respect to the parent even-even nucleus
($T_z = 4$), as a function of $4 \kappa / G$.
It is clear that when $4 \kappa / G \approx 1.4 $ proton-neutron
correlation dominates over  proton-proton and neutron-neutron
pairing correlations and the  excitation energy goes to zero. This
behavior is not a surprise and is similar to that found in
the case of the pairing plus quadrupole Hamiltonian \cite{Kis63}.
In this case, if the
quadrupole-quadrupole interaction is strong enough, the system
becomes permanently deformed. This was shown long time ago using
the "pairing plus monopole" model \cite{Aga68,Sch68} which is a
solvable two-level model obeying the algebra of SO(5). In
this letter we have applied the same ideas to the proton-neutron problem.

It must be stressed that, in realistic cases, odd-odd nuclei in the
lower energy sector of the mass parabola cannot have negative excitation
energies, as compared to the even-even nuclei with the same mass, because
it will be in contradiction with the main evidence for pairing 
effects in medium and heavy nuclei. It would
also suppress completely the double beta decay because the single
beta decay would be
allowed between even-even and odd-odd nuclei.
For the
cases where the hamiltonian (\ref{hamex}) predicts negative excitation
energies, i.e: the above mentioned overbinding of odd-odd nuclei, 
quadrupole-quadrupole
interactions and permanent deformations have to be considered.

Now we turn to the QRPA, the RQRPA and their validity.
After performing the Bogolyubov
transformations, separately, for protons and neutrons, we have
obtained the qp-hamiltonian

\begin{equation}
\begin{array}{rl}
H = &(\epsilon_p -\lambda_p) N_p + (\epsilon_n - \lambda_n) N_n +
\lambda_1 A^\dagger  A + \lambda_2 ( A^\dagger A^\dagger + A A )  \\
&-\lambda_3 ( A^\dagger B + B^\dagger A) -
\lambda_4 (A^\dagger B^\dagger + BA) + \lambda_5 B^\dagger B +
\lambda_6 (B^\dagger B^\dagger + B B) \label{hambcs}
\end{array}
\end{equation}

\noindent
being $\epsilon_p = \epsilon_n= G \Omega/2$ the quasiparticle
energies and $\lambda_p, \lambda_n$ the chemical potentials. Introducing
the quasiparticle creation operators $\alpha^\dagger_p, \alpha^\dagger_n$
\cite{Row70} and $\Omega = (2j+1)/2$, we have defined:

\begin{equation}
\begin{array}{l}
A^\dagger   = \left [ \alpha^{\dagger}_p \otimes
\alpha^{\dagger}_n \right ]^{J=0}_{M=0} ,\hspace{.5cm}
B^\dagger   = \left [ \alpha^{\dagger}_p \otimes
\alpha_{\bar n} \right ]^{J=0}_{M=0} ,\hspace{.5cm}
N_i = \sum\limits_{m_i}  \alpha^{\dagger}_{im_i}
\alpha_{im_i}\hspace{.3cm} \hbox{i=p,n} \\
\lambda_1 = 4\Omega \left [\chi (u_p^2 v_n^2 + v_p^2 u_n^2) -
\kappa (u_p^2 u_n^2 + v_p^2 v_n^2 ) \right ] ~,\hspace{.5cm}
\lambda_2 = 4\Omega ( \chi + \kappa ) u_p v_p u_n v_n ~,\\
\lambda_3 = 4\Omega ( \chi + \kappa ) u_n v_n (u_p^2-v_p^2)~,\hspace{.5cm}
\lambda_4 = 4\Omega ( \chi + \kappa ) u_p v_p (u_n^2-v_n^2) ~,\\
\lambda_5 = 4\Omega \left [\chi (u_p^2 u_n^2 + v_p^2 v_n^2) -
\kappa (u_p^2 v_n^2 + v_p^2 u_n^2 )\right ] ~,\hspace{.5cm}
\lambda_6 = - \lambda_2 ~.
\end{array}
\end{equation}

 The operators $A, A^\dagger$ together with their counterparts for
identical particles and $B, B^\dagger, N_p, N_n$ are the generators
of the SO(5) algebra \cite{Par65}.

The linearized version of hamiltonian (\ref{hambcs}) is obtained by
keeping only the first line of Eq.(5), i.e. by neglecting the scattering terms
proportional to $B$ and $B^\dagger$. Its solutions were discussed
in a previous paper \cite{Hir96a}.  

Finding the eigenvalues and eigenvectors  of hamiltonian
(\ref{hambcs}) requires the same algebraic  techniques involved in
solving the original hamiltonian. But  the  complexity of the
problem increases severely, due to the fact that neither the
quasiparticle  number nor the quasiparticle isospin projection
(or equivalently the number of proton and neutron quasiparticles)
are good quantum numbers. It implies that the dimension
of the basis increases in two orders of magnitude.
Technical details and a number
of different examples are given elsewhere \cite{Hir96b}.

The dashed line in Fig. 2 shows the dependence of the excitation
energy $E_{qp}$ for hamiltonian (\ref{hambcs}). It reproduces the
exact results reasonably up to $4 \kappa /G \approx 1.4$. From there on
it goes to zero instead of taking negative values.

 The QRPA matrix is a $ 2 \times 2$ one, with sub-matrices
$ {\cal A_{QRPA}} =  2 \epsilon + \lambda_1$ and ${\cal B_{QRPA}} = 2
\lambda_2$.
The eigen-energy is $ E_{QRPA} = [(2 \epsilon + \lambda_1)^2 - 4
\lambda_2^2]^{1/2}$. It is shown as a large-dot line in Fig. 2. It
becomes an imaginary number if $2 \lambda_2
 > 2 \epsilon + \lambda_1$.
It means that for this limit the zero-boson component of ground state
ceased to be dominant \cite{Hir96a}. The collapse occurs very near
the point where the exact excitation energies become negative. It
is a very important point.  The overestimation
of the proton-neutron correlations in the QRPA mimics the more complicated
physics found in the exact case. However it gives a clear signal
about drastic changes in the correlations, in this region of
the parameters governing the
residual pn-interaction.

In the renormalized QRPA the structure of the ground state is
included explicitly \cite{Row68}. The RQRPA energy $E_{RQRPA}$ is
always real. Its
value must be obtained by solving simultaneously a  set of
non-linear equations \cite{Toi95,Hir96a}. It is shown as a small-dot curve
in Fig. 2.
This figure strongly resemble Fig. 1 of ref. \cite{Toi95} and Fig.
2 of ref. \cite{Civ91b}, where the energy of the first excited state is
plotted
against the particle-particle strength parameter $g_{pp}$. The
curves for the QRPA and the RQRPA are quite similar to those shown
here. The new feature discussed in this letter is that the exact
excitation energies are closer to the QRPA energies rather than to
the renormalized ones.

The amplitudes $M_{2\nu}$ are evaluated as:
\begin{equation}
 M_{2\nu} ~=~ \sum_\lambda {\frac {<0_f|\beta^-| 0_\lambda >
<0_\lambda|\beta^- |0_i> } {E_\lambda -E_i + \Delta} }
\end{equation}

We have selected $\Delta = 0.5~MeV$. 
The values of $M_{2\nu}$, corresponding to the exact solution, are shown
in Fig.3 (full line) as a function of  
 $4 \kappa/ G$. For the other approximations we have 
diagonalized the hamiltonian (\ref{hambcs})  for the initial
($N_p = 6, N_n = 14$, ground state $|0_i>$) and final ($N_p =
8, N_n = 12$, ground state $|0_f>$) ground-states of the 
participant even-even nuclei. The overlap between the two descriptions of the
intermediate states belonging to the double-odd nucleus \cite{Civ87,Mut89} 
was included in the above formula (Eq.(7)).
The sum runs over the allowed intermediate states corresponding to this model
space \cite{Hir96b}.

{\bf Fig. 3}

In all the cases the curve corresponding to $M_{2\nu}$ is very similar 
to that found in realistic
calculations \cite{Vog86,Civ87,Mut89,Toi95}, including its cancellation
near the collapse of the QRPA description. The RQRPA extends this
curve far beyond the value of $\kappa$ at which the QRPA collapses.
 But the validity of this
result can be questioned because the RQRPA missed the
vanishing of the excitation energy, as we have discussed before.

To conclude, we have presented an exactly solvable model which resembles the
main features of the realistic models used to describe the structure
 of the nuclei involved in
double beta decay processes.
We have compared the exact values for the excitation energies and 
double beta decay transition amplitudes with those obtained with the
approximate qp-hamiltonian, the QRPA and renormalized
RQRPA ones.
We have shown that the collapse of the QRPA correlates with the presence of
a state with zero excitation energy.
This state is present in the exact solution of the model.
It is found that the RQRPA solutions do not show the presence of such
state. As a direct consequence of this fact this approximation gives
finite but spurious results for the transition matrix elements,
which unfortunately are not supported by the exact results.
The presence of spurious states, introduced by the renormalization
procedure, could be responsible  for such an odd behavior.
The effect of these spurious contributions upon the transition amplitudes and
the mixing of orders, in the sense of the order classfication in powers
of $1/\Omega$, found in the RQRPA wave functions are discussed in
detail somewhere else \cite{Hir96b}.

We thank J. Engel for useful comments concerning the hamiltonian
(\ref{hamex}).	O.C. thanks the hospitality of the Institute of 
Nuclear Theory of the University of Washington, where part of this work 
was performed.
Partial support of the 
Conacyt, the CONICET, and the J.S. Guggenheim Foundation is acknowledged.

\newpage

\centerline{\bf Figure Captions}

\bigskip
Fig. 1: Lowest energy states of the different nuclei, labeled
by $(T,T_Z)$, for $G = 4\kappa$. Allowed Fermi transitions are 
denoted by arrows.

\bigskip
Fig. 2: Excitation energies as a function of $4 \kappa /G$.

\bigskip
Fig. 3: $\beta\beta_{2\nu}$ transition amplitudes $M_{2\nu}$ {\em vs.}
$4 \kappa / G$.

\newpage

\setlength{\unitlength}{0.240900pt}
\ifx\plotpoint\undefined\newsavebox{\plotpoint}\fi
\sbox{\plotpoint}{\rule[-0.200pt]{0.400pt}{0.400pt}}%
\begin{picture}(1500,900)(0,0)
\font\gnuplot=cmr10 at 12pt
\gnuplot
\sbox{\plotpoint}{\rule[-0.200pt]{0.400pt}{0.400pt}}%
\put(120.0,31.0){\rule[-0.200pt]{0.400pt}{207.656pt}}
\put(120.0,31.0){\rule[-0.200pt]{4.818pt}{0.400pt}}
\put(108,31){\makebox(0,0)[r]{-8.}}
\put(1436.0,31.0){\rule[-0.200pt]{4.818pt}{0.400pt}}
\put(120.0,139.0){\rule[-0.200pt]{4.818pt}{0.400pt}}
\put(1436.0,139.0){\rule[-0.200pt]{4.818pt}{0.400pt}}
\put(120.0,247.0){\rule[-0.200pt]{4.818pt}{0.400pt}}
\put(108,247){\makebox(0,0)[r]{-7.}}
\put(1436.0,247.0){\rule[-0.200pt]{4.818pt}{0.400pt}}
\put(120.0,354.0){\rule[-0.200pt]{4.818pt}{0.400pt}}
\put(1436.0,354.0){\rule[-0.200pt]{4.818pt}{0.400pt}}
\put(120.0,462.0){\rule[-0.200pt]{4.818pt}{0.400pt}}
\put(108,462){\makebox(0,0)[r]{-6.}}
\put(1436.0,462.0){\rule[-0.200pt]{4.818pt}{0.400pt}}
\put(120.0,570.0){\rule[-0.200pt]{4.818pt}{0.400pt}}
\put(1436.0,570.0){\rule[-0.200pt]{4.818pt}{0.400pt}}
\put(120.0,678.0){\rule[-0.200pt]{4.818pt}{0.400pt}}
\put(108,678){\makebox(0,0)[r]{-5.}}
\put(1436.0,678.0){\rule[-0.200pt]{4.818pt}{0.400pt}}
\put(120.0,785.0){\rule[-0.200pt]{4.818pt}{0.400pt}}
\put(1436.0,785.0){\rule[-0.200pt]{4.818pt}{0.400pt}}
\put(120.0,893.0){\rule[-0.200pt]{4.818pt}{0.400pt}}
\put(108,893){\makebox(0,0)[r]{-4.}}
\put(1436.0,893.0){\rule[-0.200pt]{4.818pt}{0.400pt}}
\put(120.0,31.0){\rule[-0.200pt]{0.400pt}{4.818pt}}

\put(250,5){\makebox(0,0){5}}
\put(530,5){\makebox(0,0){6}}
\put(810,5){\makebox(0,0){7}}
\put(1090,5){\makebox(0,0){8}}
\put(1370,5){\makebox(0,0){9}}
\put(120.0,31.0){\rule[-0.200pt]{321.842pt}{0.400pt}}
\put(1456.0,31.0){\rule[-0.200pt]{0.400pt}{207.656pt}}
\put(120.0,893.0){\rule[-0.200pt]{321.842pt}{0.400pt}}
\put(0,780){\makebox(0,0){E [MeV]}}
\put(824,-89){\makebox(0,0){Z}}
\put(824,-500){\makebox(0,0){Figure 1}}
\put(120.0,31.0){\rule[-0.200pt]{0.400pt}{207.656pt}}
\put(147.0,742.0){\rule[-0.400pt]{51.312pt}{0.800pt}}
\put(247,780){\makebox(0,0){(6,5)}}
\put(414.0,247.0){\rule[-0.400pt]{51.553pt}{0.800pt}}
\put(514,280){\makebox(0,0){(4,4)}}
\put(681.0,440.0){\rule[-0.400pt]{51.553pt}{0.800pt}}
\put(781,473){\makebox(0,0){(4,3)}}
\put(948.0,160.0){\rule[-0.400pt]{51.553pt}{0.800pt}}
\put(1048,193){\makebox(0,0){(2,2)}}
\put(1216.0,311){\rule[-0.400pt]{51.553pt}{0.800pt}}
\put(1316,344){\makebox(0,0){(2,1)}}
\put(610,260){\vector(1,2){85}}
\put(630,355){\makebox(0,0){$t^-$}}
\put(1140,173){\vector(2,3){85}}
\put(1150,250){\makebox(0,0){$t^-$}}
\end{picture}

\clearpage
\newpage

\setlength{\unitlength}{0.240900pt}
\ifx\plotpoint\undefined\newsavebox{\plotpoint}\fi
\begin{picture}(1500,900)(0,0)
\font\gnuplot=cmr10 at 12pt
\gnuplot
\sbox{\plotpoint}{\rule[-0.200pt]{0.400pt}{0.400pt}}%
\put(120.0,318.0){\rule[-0.200pt]{321.842pt}{0.400pt}}
\put(120.0,31.0){\rule[-0.200pt]{0.400pt}{207.656pt}}
\put(120.0,31.0){\rule[-0.200pt]{4.818pt}{0.400pt}}
\put(1436.0,31.0){\rule[-0.200pt]{4.818pt}{0.400pt}}
\put(120.0,127.0){\rule[-0.200pt]{4.818pt}{0.400pt}}
\put(108,127){\makebox(0,0)[r]{-1}}
\put(1436.0,127.0){\rule[-0.200pt]{4.818pt}{0.400pt}}
\put(120.0,223.0){\rule[-0.200pt]{4.818pt}{0.400pt}}
\put(1436.0,223.0){\rule[-0.200pt]{4.818pt}{0.400pt}}
\put(120.0,318.0){\rule[-0.200pt]{4.818pt}{0.400pt}}
\put(108,318){\makebox(0,0)[r]{0}}
\put(1436.0,318.0){\rule[-0.200pt]{4.818pt}{0.400pt}}
\put(120.0,414.0){\rule[-0.200pt]{4.818pt}{0.400pt}}
\put(1436.0,414.0){\rule[-0.200pt]{4.818pt}{0.400pt}}
\put(120.0,510.0){\rule[-0.200pt]{4.818pt}{0.400pt}}
\put(108,510){\makebox(0,0)[r]{1}}
\put(1436.0,510.0){\rule[-0.200pt]{4.818pt}{0.400pt}}
\put(120.0,606.0){\rule[-0.200pt]{4.818pt}{0.400pt}}
\put(1436.0,606.0){\rule[-0.200pt]{4.818pt}{0.400pt}}
\put(120.0,701.0){\rule[-0.200pt]{4.818pt}{0.400pt}}
\put(108,701){\makebox(0,0)[r]{2}}
\put(1436.0,701.0){\rule[-0.200pt]{4.818pt}{0.400pt}}
\put(120.0,797.0){\rule[-0.200pt]{4.818pt}{0.400pt}}
\put(1436.0,797.0){\rule[-0.200pt]{4.818pt}{0.400pt}}
\put(120.0,893.0){\rule[-0.200pt]{4.818pt}{0.400pt}}
\put(108,893){\makebox(0,0)[r]{3}}
\put(1436.0,893.0){\rule[-0.200pt]{4.818pt}{0.400pt}}
\put(120.0,31.0){\rule[-0.200pt]{0.400pt}{4.818pt}}
\put(120,5){\makebox(0,0){0}}
\put(120.0,873.0){\rule[-0.200pt]{0.400pt}{4.818pt}}
\put(254.0,31.0){\rule[-0.200pt]{0.400pt}{4.818pt}}
\put(254,5){\makebox(0,0){0.2}}
\put(254.0,873.0){\rule[-0.200pt]{0.400pt}{4.818pt}}
\put(387.0,31.0){\rule[-0.200pt]{0.400pt}{4.818pt}}
\put(387,5){\makebox(0,0){0.4}}
\put(387.0,873.0){\rule[-0.200pt]{0.400pt}{4.818pt}}
\put(521.0,31.0){\rule[-0.200pt]{0.400pt}{4.818pt}}
\put(521,5){\makebox(0,0){0.6}}
\put(521.0,873.0){\rule[-0.200pt]{0.400pt}{4.818pt}}
\put(654.0,31.0){\rule[-0.200pt]{0.400pt}{4.818pt}}
\put(654,5){\makebox(0,0){0.8}}
\put(654.0,873.0){\rule[-0.200pt]{0.400pt}{4.818pt}}
\put(788.0,31.0){\rule[-0.200pt]{0.400pt}{4.818pt}}
\put(788,5){\makebox(0,0){1}}
\put(788.0,873.0){\rule[-0.200pt]{0.400pt}{4.818pt}}
\put(922.0,31.0){\rule[-0.200pt]{0.400pt}{4.818pt}}
\put(922,5){\makebox(0,0){1.2}}
\put(922.0,873.0){\rule[-0.200pt]{0.400pt}{4.818pt}}
\put(1055.0,31.0){\rule[-0.200pt]{0.400pt}{4.818pt}}
\put(1055,5){\makebox(0,0){1.4}}
\put(1055.0,873.0){\rule[-0.200pt]{0.400pt}{4.818pt}}
\put(1189.0,31.0){\rule[-0.200pt]{0.400pt}{4.818pt}}
\put(1189,5){\makebox(0,0){1.6}}
\put(1189.0,873.0){\rule[-0.200pt]{0.400pt}{4.818pt}}
\put(1322.0,31.0){\rule[-0.200pt]{0.400pt}{4.818pt}}
\put(1322,5){\makebox(0,0){1.8}}
\put(1322.0,873.0){\rule[-0.200pt]{0.400pt}{4.818pt}}
\put(1456.0,31.0){\rule[-0.200pt]{0.400pt}{4.818pt}}
\put(1456,5){\makebox(0,0){2}}
\put(1456.0,873.0){\rule[-0.200pt]{0.400pt}{4.818pt}}
\put(120.0,31.0){\rule[-0.200pt]{321.842pt}{0.400pt}}
\put(1456.0,31.0){\rule[-0.200pt]{0.400pt}{207.656pt}}
\put(120.0,893.0){\rule[-0.200pt]{321.842pt}{0.400pt}}
\put(12,790){\makebox(0,0){E [MeV]}}
\put(788,-90){\makebox(0,0){$4 \kappa / G$}}
\put(788,-320){\makebox(0,0){Figure 2}}
\put(120.0,31.0){\rule[-0.200pt]{0.400pt}{207.656pt}}
\sbox{\plotpoint}{\rule[-0.400pt]{0.800pt}{0.800pt}}%
\put(1376,850){\makebox(0,0)[r]{exact}}
\put(1388.0,850.0){\rule[-0.400pt]{8.672pt}{0.800pt}}
\put(120,807){\usebox{\plotpoint}}
\multiput(120.00,805.09)(1.487,-0.505){39}{\rule{2.530pt}{0.122pt}}
\multiput(120.00,805.34)(61.748,-23.000){2}{\rule{1.265pt}{0.800pt}}
\multiput(187.00,782.09)(1.422,-0.504){41}{\rule{2.433pt}{0.122pt}}
\multiput(187.00,782.34)(61.949,-24.000){2}{\rule{1.217pt}{0.800pt}}
\multiput(254.00,758.09)(1.290,-0.504){45}{\rule{2.231pt}{0.121pt}}
\multiput(254.00,758.34)(61.370,-26.000){2}{\rule{1.115pt}{0.800pt}}
\multiput(320.00,732.09)(1.259,-0.504){47}{\rule{2.185pt}{0.121pt}}
\multiput(320.00,732.34)(62.465,-27.000){2}{\rule{1.093pt}{0.800pt}}
\multiput(387.00,705.09)(1.213,-0.504){49}{\rule{2.114pt}{0.121pt}}
\multiput(387.00,705.34)(62.612,-28.000){2}{\rule{1.057pt}{0.800pt}}
\multiput(454.00,677.09)(1.092,-0.503){55}{\rule{1.929pt}{0.121pt}}
\multiput(454.00,677.34)(62.996,-31.000){2}{\rule{0.965pt}{0.800pt}}
\multiput(521.00,646.09)(1.057,-0.503){57}{\rule{1.875pt}{0.121pt}}
\multiput(521.00,646.34)(63.108,-32.000){2}{\rule{0.938pt}{0.800pt}}
\multiput(588.00,614.09)(0.979,-0.503){61}{\rule{1.753pt}{0.121pt}}
\multiput(588.00,614.34)(62.362,-34.000){2}{\rule{0.876pt}{0.800pt}}
\multiput(654.00,580.09)(0.937,-0.503){65}{\rule{1.689pt}{0.121pt}}
\multiput(654.00,580.34)(63.495,-36.000){2}{\rule{0.844pt}{0.800pt}}
\multiput(721.00,544.09)(0.887,-0.503){69}{\rule{1.611pt}{0.121pt}}
\multiput(721.00,544.34)(63.657,-38.000){2}{\rule{0.805pt}{0.800pt}}
\multiput(788.00,506.09)(0.842,-0.502){73}{\rule{1.540pt}{0.121pt}}
\multiput(788.00,506.34)(63.804,-40.000){2}{\rule{0.770pt}{0.800pt}}
\multiput(855.00,466.09)(0.782,-0.502){79}{\rule{1.447pt}{0.121pt}}
\multiput(855.00,466.34)(63.998,-43.000){2}{\rule{0.723pt}{0.800pt}}
\multiput(922.00,423.09)(0.752,-0.502){81}{\rule{1.400pt}{0.121pt}}
\multiput(922.00,423.34)(63.094,-44.000){2}{\rule{0.700pt}{0.800pt}}
\multiput(988.00,379.09)(0.747,-0.502){83}{\rule{1.391pt}{0.121pt}}
\multiput(988.00,379.34)(64.113,-45.000){2}{\rule{0.696pt}{0.800pt}}
\multiput(1055.00,334.09)(0.730,-0.502){85}{\rule{1.365pt}{0.121pt}}
\multiput(1055.00,334.34)(64.166,-46.000){2}{\rule{0.683pt}{0.800pt}}
\multiput(1122.00,288.09)(0.714,-0.502){87}{\rule{1.340pt}{0.121pt}}
\multiput(1122.00,288.34)(64.218,-47.000){2}{\rule{0.670pt}{0.800pt}}
\multiput(1189.00,241.09)(0.747,-0.502){83}{\rule{1.391pt}{0.121pt}}
\multiput(1189.00,241.34)(64.113,-45.000){2}{\rule{0.696pt}{0.800pt}}
\multiput(1256.00,196.09)(0.735,-0.502){83}{\rule{1.373pt}{0.121pt}}
\multiput(1256.00,196.34)(63.150,-45.000){2}{\rule{0.687pt}{0.800pt}}
\multiput(1322.00,151.09)(0.782,-0.502){79}{\rule{1.447pt}{0.121pt}}
\multiput(1322.00,151.34)(63.998,-43.000){2}{\rule{0.723pt}{0.800pt}}
\multiput(1389.00,108.09)(0.801,-0.502){77}{\rule{1.476pt}{0.121pt}}
\multiput(1389.00,108.34)(63.936,-42.000){2}{\rule{0.738pt}{0.800pt}}
\sbox{\plotpoint}{\rule[-0.200pt]{0.400pt}{0.400pt}}%
\put(1376,775){\makebox(0,0)[r]{qp-hamiltonian}}
\put(1388.0,775.0){\rule[-0.200pt]{8.672pt}{0.400pt}}
\put(120,814){\usebox{\plotpoint}}
\multiput(120.00,812.92)(1.613,-0.496){30}{\rule{1.376pt}{0.119pt}}
\multiput(187.00,791.92)(1.470,-0.496){30}{\rule{1.265pt}{0.120pt}}
\multiput(254.00,768.92)(1.448,-0.496){30}{\rule{1.248pt}{0.120pt}}
\multiput(320.00,745.92)(1.408,-0.496){33}{\rule{1.217pt}{0.120pt}}
\multiput(387.00,721.92)(1.298,-0.497){33}{\rule{1.131pt}{0.120pt}}
\multiput(454.00,695.92)(1.249,-0.497){35}{\rule{1.093pt}{0.120pt}}
\multiput(521.00,668.92)(1.162,-0.497){40}{\rule{1.024pt}{0.120pt}}
\multiput(588.00,639.92)(1.036,-0.497){45}{\rule{0.925pt}{0.120pt}}
\multiput(654.00,607.92)(0.989,-0.498){50}{\rule{0.888pt}{0.120pt}}
\multiput(721.00,573.92)(0.884,-0.498){55}{\rule{0.805pt}{0.120pt}}
\multiput(788.00,535.92)(0.799,-0.498){60}{\rule{0.738pt}{0.120pt}}
\multiput(855.00,493.92)(0.745,-0.498){65}{\rule{0.696pt}{0.120pt}}
\multiput(922.00,448.92)(0.703,-0.498){70}{\rule{0.662pt}{0.120pt}}
\multiput(988.00,401.92)(0.819,-0.498){55}{\rule{0.754pt}{0.120pt}}
\multiput(1055.00,360.92)(1.351,-0.497){33}{\rule{1.172pt}{0.120pt}}
\multiput(1122.00,335.92)(2.866,-0.492){12}{\rule{2.333pt}{0.119pt}}
\multiput(1189.00,323.94)(9.693,-0.468){3}{\rule{6.800pt}{0.113pt}}
\put(1256,319.17){\rule{13.300pt}{0.400pt}}
\put(1322.0,319.0){\rule[-0.200pt]{32.281pt}{0.400pt}}
\sbox{\plotpoint}{\rule[-0.500pt]{2.00pt}{2.00pt}}%
\put(1376,700){\makebox(0,0)[r]{QRPA}}
\multiput(1388,700)(20.756,0.000){2}{\usebox{\plotpoint}}
\put(1424,700){\usebox{\plotpoint}}
\put(120,806){\usebox{\plotpoint}}
\multiput(120,806)(19.968,-5.663){4}{\usebox{\plotpoint}}
\multiput(187,787)(19.805,-6.208){3}{\usebox{\plotpoint}}
\multiput(254,766)(19.778,-6.293){4}{\usebox{\plotpoint}}
\multiput(320,745)(19.720,-6.475){3}{\usebox{\plotpoint}}
\multiput(387,723)(19.631,-6.739){3}{\usebox{\plotpoint}}
\multiput(454,700)(19.350,-7.509){4}{\usebox{\plotpoint}}
\multiput(521,674)(19.251,-7.758){3}{\usebox{\plotpoint}}
\multiput(588,647)(19.002,-8.349){4}{\usebox{\plotpoint}}
\multiput(654,618)(18.620,-9.171){3}{\usebox{\plotpoint}}
\multiput(721,585)(18.169,-10.034){4}{\usebox{\plotpoint}}
\multiput(788,548)(17.349,-11.393){4}{\usebox{\plotpoint}}
\multiput(855,504)(15.462,-13.846){4}{\usebox{\plotpoint}}
\multiput(922,444)(9.631,-18.386){7}{\usebox{\plotpoint}}
\put(1456,318){\usebox{\plotpoint}}
\sbox{\plotpoint}{\rule[-0.200pt]{1.000pt}{1.000pt}}%
\put(1376,625){\makebox(0,0)[r]{RQRPA}}
\multiput(1388,625)(20.756,0.000){2}{\usebox{\plotpoint}}
\put(1424,625){\usebox{\plotpoint}}
\put(120,806){\usebox{\plotpoint}}
\multiput(120,806)(19.968,-5.663){4}{\usebox{\plotpoint}}
\multiput(187,787)(19.805,-6.208){3}{\usebox{\plotpoint}}
\multiput(254,766)(19.778,-6.293){4}{\usebox{\plotpoint}}
\multiput(320,745)(19.720,-6.475){3}{\usebox{\plotpoint}}
\multiput(387,723)(19.631,-6.739){3}{\usebox{\plotpoint}}
\multiput(454,700)(19.446,-7.256){4}{\usebox{\plotpoint}}
\multiput(521,675)(19.251,-7.758){3}{\usebox{\plotpoint}}
\multiput(588,648)(19.107,-8.106){4}{\usebox{\plotpoint}}
\multiput(654,620)(18.837,-8.716){3}{\usebox{\plotpoint}}
\multiput(721,589)(18.509,-9.392){4}{\usebox{\plotpoint}}
\multiput(788,555)(18.054,-10.240){4}{\usebox{\plotpoint}}
\multiput(855,517)(17.586,-11.024){3}{\usebox{\plotpoint}}
\multiput(922,475)(17.511,-11.143){4}{\usebox{\plotpoint}}
\multiput(988,433)(18.283,-9.824){4}{\usebox{\plotpoint}}
\multiput(1055,397)(19.631,-6.739){3}{\usebox{\plotpoint}}
\multiput(1122,374)(20.317,-4.245){4}{\usebox{\plotpoint}}
\multiput(1189,360)(20.571,-2.763){3}{\usebox{\plotpoint}}
\multiput(1256,351)(20.670,-1.879){3}{\usebox{\plotpoint}}
\multiput(1322,345)(20.719,-1.237){3}{\usebox{\plotpoint}}
\multiput(1389,341)(20.735,-0.928){4}{\usebox{\plotpoint}}
\put(1456,338){\usebox{\plotpoint}}
\end{picture}

\clearpage
\newpage

\setlength{\unitlength}{0.240900pt}
\ifx\plotpoint\undefined\newsavebox{\plotpoint}\fi
\sbox{\plotpoint}{\rule[-0.200pt]{0.400pt}{0.400pt}}%
\begin{picture}(1500,900)(0,0)
\font\gnuplot=cmr10 at 12pt
\gnuplot
\sbox{\plotpoint}{\rule[-0.200pt]{0.400pt}{0.400pt}}%
\put(120.0,318.0){\rule[-0.200pt]{321.842pt}{0.400pt}}
\put(120.0,31.0){\rule[-0.200pt]{0.400pt}{207.656pt}}
\put(120.0,31.0){\rule[-0.200pt]{4.818pt}{0.400pt}}
\put(108,31){\makebox(0,0)[r]{-1}}
\put(1436.0,31.0){\rule[-0.200pt]{4.818pt}{0.400pt}}
\put(120.0,175.0){\rule[-0.200pt]{4.818pt}{0.400pt}}
\put(1436.0,175.0){\rule[-0.200pt]{4.818pt}{0.400pt}}
\put(120.0,318.0){\rule[-0.200pt]{4.818pt}{0.400pt}}
\put(108,318){\makebox(0,0)[r]{0}}
\put(1436.0,318.0){\rule[-0.200pt]{4.818pt}{0.400pt}}
\put(120.0,462.0){\rule[-0.200pt]{4.818pt}{0.400pt}}
\put(1436.0,462.0){\rule[-0.200pt]{4.818pt}{0.400pt}}
\put(120.0,606.0){\rule[-0.200pt]{4.818pt}{0.400pt}}
\put(108,606){\makebox(0,0)[r]{1}}
\put(1436.0,606.0){\rule[-0.200pt]{4.818pt}{0.400pt}}
\put(120.0,749.0){\rule[-0.200pt]{4.818pt}{0.400pt}}
\put(1436.0,749.0){\rule[-0.200pt]{4.818pt}{0.400pt}}
\put(120.0,893.0){\rule[-0.200pt]{4.818pt}{0.400pt}}
\put(108,893){\makebox(0,0)[r]{2}}
\put(1436.0,893.0){\rule[-0.200pt]{4.818pt}{0.400pt}}
\put(120.0,31.0){\rule[-0.200pt]{0.400pt}{4.818pt}}
\put(120,5){\makebox(0,0){0}}
\put(120.0,873.0){\rule[-0.200pt]{0.400pt}{4.818pt}}
\put(311.0,31.0){\rule[-0.200pt]{0.400pt}{4.818pt}}
\put(311,5){\makebox(0,0){0.2}}
\put(311.0,873.0){\rule[-0.200pt]{0.400pt}{4.818pt}}
\put(502.0,31.0){\rule[-0.200pt]{0.400pt}{4.818pt}}
\put(502,5){\makebox(0,0){0.4}}
\put(502.0,873.0){\rule[-0.200pt]{0.400pt}{4.818pt}}
\put(693.0,31.0){\rule[-0.200pt]{0.400pt}{4.818pt}}
\put(693,5){\makebox(0,0){0.6}}
\put(693.0,873.0){\rule[-0.200pt]{0.400pt}{4.818pt}}
\put(883.0,31.0){\rule[-0.200pt]{0.400pt}{4.818pt}}
\put(883,5){\makebox(0,0){0.8}}
\put(883.0,873.0){\rule[-0.200pt]{0.400pt}{4.818pt}}
\put(1074.0,31.0){\rule[-0.200pt]{0.400pt}{4.818pt}}
\put(1074,5){\makebox(0,0){1}}
\put(1074.0,873.0){\rule[-0.200pt]{0.400pt}{4.818pt}}
\put(1265.0,31.0){\rule[-0.200pt]{0.400pt}{4.818pt}}
\put(1265,5){\makebox(0,0){1.2}}
\put(1265.0,873.0){\rule[-0.200pt]{0.400pt}{4.818pt}}
\put(1456.0,31.0){\rule[-0.200pt]{0.400pt}{4.818pt}}
\put(1456,5){\makebox(0,0){1.4}}
\put(1456.0,873.0){\rule[-0.200pt]{0.400pt}{4.818pt}}
\put(120.0,31.0){\rule[-0.200pt]{321.842pt}{0.400pt}}
\put(1456.0,31.0){\rule[-0.200pt]{0.400pt}{207.656pt}}
\put(120.0,893.0){\rule[-0.200pt]{321.842pt}{0.400pt}}
\put(-40,749){\makebox(0,0){$M_{2\nu} [MeV^{-1}]$}}
\put(788,-90){\makebox(0,0){$4 \kappa / G$}}
\put(788,-290){\makebox(0,0){Figure 3}}
\put(120.0,31.0){\rule[-0.200pt]{0.400pt}{207.656pt}}
\sbox{\plotpoint}{\rule[-0.400pt]{0.800pt}{0.800pt}}%
\put(1376,850){\makebox(0,0)[r]{exact}}
\put(1388.0,850.0){\rule[-0.400pt]{8.672pt}{0.800pt}}
\put(120,743){\usebox{\plotpoint}}
\multiput(120.00,741.08)(6.842,-0.520){9}{\rule{9.700pt}{0.125pt}}
\multiput(120.00,741.34)(74.867,-8.000){2}{\rule{4.850pt}{0.800pt}}
\multiput(215.00,733.08)(5.975,-0.516){11}{\rule{8.733pt}{0.124pt}}
\multiput(215.00,733.34)(77.874,-9.000){2}{\rule{4.367pt}{0.800pt}}
\multiput(311.00,724.08)(4.678,-0.512){15}{\rule{7.109pt}{0.123pt}}
\multiput(311.00,724.34)(80.245,-11.000){2}{\rule{3.555pt}{0.800pt}}
\multiput(406.00,713.09)(3.360,-0.508){23}{\rule{5.320pt}{0.122pt}}
\multiput(406.00,713.34)(84.958,-15.000){2}{\rule{2.660pt}{0.800pt}}
\multiput(502.00,698.09)(2.586,-0.506){31}{\rule{4.200pt}{0.122pt}}
\multiput(502.00,698.34)(86.283,-19.000){2}{\rule{2.100pt}{0.800pt}}
\multiput(597.00,679.09)(1.812,-0.504){47}{\rule{3.044pt}{0.121pt}}
\multiput(597.00,679.34)(89.681,-27.000){2}{\rule{1.522pt}{0.800pt}}
\multiput(693.00,652.09)(1.263,-0.503){69}{\rule{2.200pt}{0.121pt}}
\multiput(693.00,652.34)(90.434,-38.000){2}{\rule{1.100pt}{0.800pt}}
\multiput(788.00,614.09)(0.836,-0.502){107}{\rule{1.533pt}{0.121pt}}
\multiput(788.00,614.34)(91.817,-57.000){2}{\rule{0.767pt}{0.800pt}}
\multiput(883.00,557.09)(0.527,-0.501){175}{\rule{1.044pt}{0.121pt}}
\multiput(883.00,557.34)(93.833,-91.000){2}{\rule{0.522pt}{0.800pt}}
\multiput(980.41,461.93)(0.501,-0.791){183}{\rule{0.121pt}{1.463pt}}
\multiput(977.34,464.96)(95.000,-146.963){2}{\rule{0.800pt}{0.732pt}}
\multiput(1075.41,308.35)(0.501,-1.333){185}{\rule{0.121pt}{2.325pt}}
\multiput(1072.34,313.17)(96.000,-250.174){2}{\rule{0.800pt}{1.163pt}}
\multiput(1171.40,46.99)(0.526,-2.665){7}{\rule{0.127pt}{3.857pt}}
\multiput(1168.34,54.99)(7.000,-23.994){2}{\rule{0.800pt}{1.929pt}}
\sbox{\plotpoint}{\rule[-0.200pt]{0.400pt}{0.400pt}}%
\put(1376,775){\makebox(0,0)[r]{qp-hamiltonian}}
\put(1388.0,775.0){\rule[-0.200pt]{8.672pt}{0.400pt}}
\put(120,760){\usebox{\plotpoint}}
\multiput(120.00,758.95)(21.002,-0.447){2}{\rule{12.767pt}{0.108pt}}
\multiput(215.00,755.93)(10.617,-0.477){4}{\rule{7.780pt}{0.115pt}}
\multiput(311.00,750.93)(8.537,-0.482){6}{\rule{6.433pt}{0.116pt}}
\multiput(406.00,744.93)(6.305,-0.488){8}{\rule{4.900pt}{0.117pt}}
\multiput(502.00,736.92)(4.458,-0.492){12}{\rule{3.555pt}{0.118pt}}
\multiput(597.00,725.92)(2.874,-0.495){20}{\rule{2.359pt}{0.119pt}}
\multiput(693.00,708.92)(2.000,-0.496){30}{\rule{1.683pt}{0.120pt}}
\multiput(788.00,684.92)(1.290,-0.498){45}{\rule{1.127pt}{0.120pt}}
\multiput(883.00,647.92)(0.829,-0.499){70}{\rule{0.762pt}{0.120pt}}
\multiput(979.58,588.85)(0.499,-0.521){120}{\rule{0.120pt}{0.517pt}}
\multiput(1074.58,488.42)(0.499,-0.954){120}{\rule{0.120pt}{0.863pt}}
\multiput(1170.58,302.26)(0.499,-1.910){90}{\rule{0.120pt}{1.623pt}}
\sbox{\plotpoint}{\rule[-0.500pt]{2.000pt}{2.000pt}}%
\put(1376,700){\makebox(0,0)[r]{QRPA}}
\multiput(1388,700)(20.756,0.000){2}{\usebox{\plotpoint}}
\put(1424,700){\usebox{\plotpoint}}
\put(120,763){\usebox{\plotpoint}}
\multiput(120,763)(20.737,-0.873){5}{\usebox{\plotpoint}}
\multiput(215,759)(20.715,-1.295){5}{\usebox{\plotpoint}}
\multiput(311,753)(20.682,-1.742){4}{\usebox{\plotpoint}}
\multiput(406,745)(20.665,-1.937){5}{\usebox{\plotpoint}}
\multiput(502,736)(20.564,-2.814){5}{\usebox{\plotpoint}}
\multiput(597,723)(20.438,-3.619){4}{\usebox{\plotpoint}}
\multiput(693,706)(20.019,-5.479){5}{\usebox{\plotpoint}}
\multiput(788,680)(18.909,-8.559){5}{\usebox{\plotpoint}}
\multiput(883,637)(15.863,-13.385){6}{\usebox{\plotpoint}}
\multiput(979,556)(7.694,-19.277){12}{\usebox{\plotpoint}}
\put(1361,318){\usebox{\plotpoint}}
\sbox{\plotpoint}{\rule[-0.200pt]{1.000pt}{1.000pt}}%
\put(1376,625){\makebox(0,0)[r]{RQRPA}}
\multiput(1388,625)(20.756,0.000){2}{\usebox{\plotpoint}}
\put(1424,625){\usebox{\plotpoint}}
\put(120,763){\usebox{\plotpoint}}
\multiput(120,763)(20.727,-1.091){5}{\usebox{\plotpoint}}
\multiput(215,758)(20.715,-1.295){5}{\usebox{\plotpoint}}
\multiput(311,752)(20.699,-1.525){4}{\usebox{\plotpoint}}
\multiput(406,745)(20.644,-2.150){5}{\usebox{\plotpoint}}
\multiput(502,735)(20.564,-2.814){5}{\usebox{\plotpoint}}
\multiput(597,722)(20.438,-3.619){4}{\usebox{\plotpoint}}
\multiput(693,705)(20.123,-5.084){5}{\usebox{\plotpoint}}
\multiput(788,681)(19.271,-7.708){5}{\usebox{\plotpoint}}
\multiput(883,643)(17.518,-11.131){5}{\usebox{\plotpoint}}
\multiput(979,582)(13.566,-15.708){7}{\usebox{\plotpoint}}
\multiput(1074,472)(8.838,-18.780){11}{\usebox{\plotpoint}}
\multiput(1170,268)(5.879,-19.905){12}{\usebox{\plotpoint}}
\put(1240,31){\usebox{\plotpoint}}
\end{picture}

\end{document}